\documentclass[twocolumn,notitlepage,superscriptaddress,nogroupedaddress]{revtex4-2}

\usepackage{ifpdf}
\usepackage{graphicx}

\usepackage{amsmath}
\usepackage{algorithmic}
\usepackage{array}

\newcommand\id{\mathbb{I}}
\newcommand\ii{\mathcal{I}}
\newcommand\jj{\mathcal{J}}

\usepackage{amsfonts}
\usepackage{amssymb}
\usepackage{float}
\usepackage{hyperref}
\usepackage{color}
\usepackage{graphicx}
\usepackage{caption}
\usepackage{subcaption}
\usepackage[utf8]{inputenc}
\usepackage{url} 
\usepackage{mathtools}
\usepackage{amsmath}
\usepackage{geometry}

\begin{document}

\title{Spectral Pruning of Fully Connected Layers: Ranking the Nodes Based on the Eigenvalues}

\author{Lorenzo Buffoni}
\affiliation{Physics of Information and Quantum Technologies Group, Instituto de Telecomunicaç\~oes, Lisbon, Portugal}
\affiliation{CSDC, Department of Physics and Astronomy, University of Florence, Sesto Fiorentino, Italy}

\author{Enrico Civitelli}
\affiliation{LabGOL, Department of Information Engineering, University of Florence, Florence, Italy}

\author{Lorenzo Giambagli}
\affiliation{CSDC, Department of Physics and Astronomy, University of Florence, Sesto Fiorentino, Italy}

\author{Lorenzo Chicchi}
\affiliation{CSDC, Department of Physics and Astronomy, University of Florence, Sesto Fiorentino, Italy}

\author{Duccio Fanelli}
\affiliation{CSDC, Department of Physics and Astronomy, University of Florence, Sesto Fiorentino, Italy}

\begin{abstract}
	Training of neural networks can be reformulated in spectral space, by allowing eigenvalues and eigenvectors of the network to act as target of the optimization instead of the individual weights. Working in this setting, we show that the eigenvalues can be used to rank the nodes' importance within the ensemble. Indeed, we will prove that sorting the nodes based on their associated eigenvalues, enables effective pre- and post-processing pruning strategies to yield massively compacted networks (in terms of the number of composing neurons) with virtually unchanged performance. The proposed methods are tested for different architectures, with just a single or multiple hidden layers, and against distinct classification tasks of general interest.
\end{abstract}

\maketitle

\section{Introduction}
Automated learning via deep neural networks is gaining increasing popularity, as a ductile procedure to address a widespread plethora of interdisciplinary applications \cite{he2018amc, sutton2018reinforcement, grigorescu2020survey}. In standard neural network training one seeks to optimise the weights that link pairs of neurons belonging to adjacent layers of the selected architecture \cite{Goodfellow-et-al-2016}. This is achieved by computing the gradient of the loss with respect to the sought weights, a procedure which amounts to operate in the so called direct space of the network \cite{spec_learn}. Alternatively, the learning can be carried out in reciprocal space: the spectral attributes (eigenvalues and eigenvectors) of the transfer operators that underlie information handling across layers define the actual target of the optimisation. This procedure, first introduced in \cite{spec_learn} and further refined in \cite{chicchi2021training}, enables a substantial compression of the space of trainable parameters. The spectral method leverages on a limited subset of key parameters which impact on the whole set of weights in direct space. Particularly relevant, in this respect, is the setting where the eigenmodes of the inter-layer transfer operators align along random directions. In this case, the associated eigenvalues constitute the sole trainable parameters. When employed for classifications tasks, the accuracy displayed by the spectral scheme restricted to operate with eigenvalues is slightly worse than that reported when the learning is carried in direct space, for an identical architecture and by employing the full set of trainable parameters. To bridge the gap between conventional and spectral methods in terms of measured performances, one can also train the elements that populate the non trivial block of the eigenvectors matrix \cite{spec_learn}. By resorting to apt decomposition schemes, it is still possible to contain the total number of trainable parameters, while reaching stunning performances in terms of classification outcomes \cite{chicchi2021training}.   

In this paper we will discuss a relevant byproduct of the spectral learning scheme. More specifically, we will argue that the eigenvalues do provide a reliable ranking of the nodes, in terms of their associated contribution to the overall performance of the trained network. Working along these lines, we will empirically prove that the absolute value of the eigenvalues is an excellent marker of the node's significance in carrying out the assigned discrimination task. This observation can be effectively exploited, downstream of training, to filter the nodes in terms of their relative importance and prune the unessential units so as to yield a more compact model, with almost identical classification abilities. The effectiveness of the proposed method has been tested for different feed-forward architectures, with just a single or multiple hidden layers, by invoking several activation functions, and against distinct datasets for image recognition, with various levels of inherent complexity. Building on these findings, we will also propose a two stages training protocol to generate minimal networks (in terms of allowed computing neurons) which outperform those obtained by hacking off dispensable units from a large, fully trained, apparatus. This is a viable strategy to discover a ``winning ticket'' \cite{frankle2018lottery}: dense (randomly-initialized) feed-forward networks contain sub-networks (aka winning tickets) with recorded performance comparable to those displayed by their unaltered homologues, after a proper round of training.

The paper is organized as follows. In the next section we will discuss the mathematical foundation and set the notation of the spectral learning scheme. We will then move on to illustrating the results of the proposed spectral pruning strategy, after a short account of the alternative methods available in the literature. Finally, we will sum up and draw our conclusions.  
The details about the proposed schemes are discussed in the Methods Section.

\section{Spectral approach to learning}
This Section is devoted to reviewing the spectral approach to the training of deep neural networks. The discussion will follow mainly \cite{chicchi2021training}, where an extension of the method originally introduced in \cite{spec_learn} is handed over.

Consider a deep feed-forward network made of $\ell$ distinct layers. Each layer is labelled with a discrete index $i$ $(=1,...,\ell)$. Denote by  $N_i$ the number of the neurons, the individual computing units, that pertain to layer $i$. Then,
we posit $N=\sum_{i=1}^{\ell} N_i$ and introduce a column vector $\vec{x}^{(1)}$, of size $N$, the first $N_1$ entries referring to the supplied input signal. As anticipated, we will be mainly concerned with datasets for image recognition, so we will use this specific case to illustrate the more general approach of spectral learning. This means that, the first $N_1$ elements of $\vec{x}^{(1)}$ are the intensities (from the top-left to the bottom-right, moving horizontally) as displayed on the pixels of the image presented as an input.  All other entries of $\vec{x}^{(1)}$ are identically equal to zero. 

\begin{figure}[!ht]
	\centering
	\includegraphics[width = 0.4\textwidth]{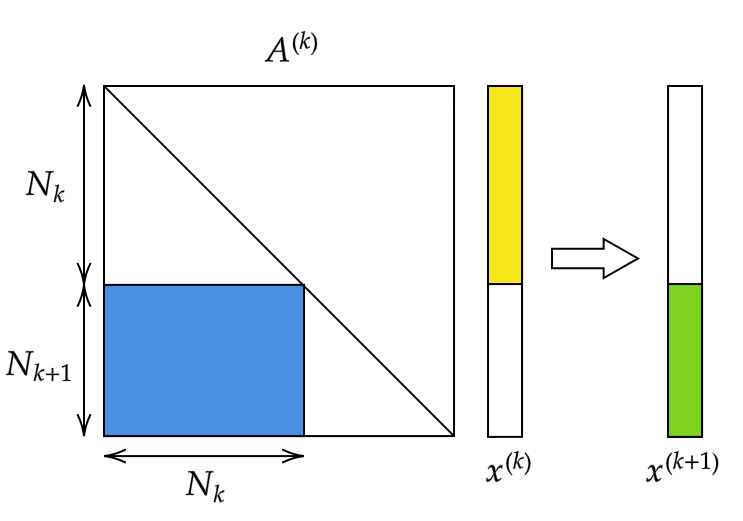}
	\caption{A schematic outline of the structure of transfer matrix ${\mathbf A}^{(k)}$, bridging layer $k$ to layer $k+1$. The action of ${\mathbf A}^{(k)}$ on $\vec{x}^{(k)}$ is also graphically illustrated.}
	\label{f:Lin_transf}
\end{figure}

\begin{figure}[!ht]
	\centering
	\includegraphics[width = 0.4\textwidth]{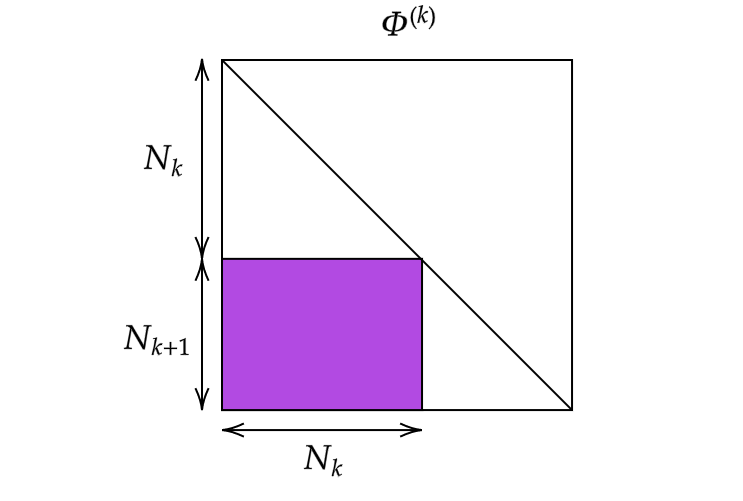}
	\caption{The structure of matrix ${\mathbf \Phi}^{(k)}$ is schematically displayed.}
	\label{f:base}
\end{figure}

The aim of the procedure is to map $\vec{x}^{(1)}$ into an output vector $\vec{x}^{(\ell)}$ , still of  size $N$: the last  $N_{\ell}$ elements are the intensities displayed at the output nodes, where reading is eventually performed. The applied transformation is composed by a suite of linear operations, interposed to non linear filters. To exemplify the overall strategy, consider the generic vector $\vec{x}^{(k)}$, with $k=1,..., \ell-1$, as obtained after $k$ execution of the above procedure. At the successive iteration, one gets $\vec{x}^{(k+1)}= {\mathbf A}^{(k)} \vec{x}_{(k)}$, where ${\mathbf A}^{(k)}$ is a $N \times N$ matrix with a rather specific structure, as elucidated in the following and schematically depicted in Fig. \ref{f:Lin_transf}. Further, a suitably defined non-linear function $f(\cdot, \beta_k)$ is applied to $\vec{x}^{(k+1)}$, where $\beta_k$ identifies an optional bias. To proceed in the analysis, we cast ${\mathbf A}^{(k)}={\mathbf \Phi}^{(k)} {\mathbf \Lambda}^{(k)} \left({\mathbf \Phi}^{(k)}\right)^{-1}$ by invoking spectral decomposition. Here,  ${\mathbf \Lambda}^{(k)}$ denotes the diagonal matrix of the eigenvalues of ${\mathbf A}^{(k)}$. Following \cite{chicchi2021training}, we set $\left({\mathbf \Lambda}^{(k)} \right)_{jj} = 1$ for $j< \sum_{i=1}^{k-1} N_i$ and $j> \sum_{i=1}^{k+1} N_i$. The remaining $N_k+N_{k+1}$ elements are initially assigned to random entries, as e.g. extracted from a uniform distribution, and define a first basin of target variables for the spectral learning scheme. Then,  ${\mathbf \Phi}^{(k)}$ is the identity matrix  $\id{}_{N \times N}$, with the inclusion of a sub-diagonal $N_{k+1} \times N_{k}$ block, denoted by {\boldmath$\phi$}$^{(k)}$, see Fig. \ref{f:base}. This choice amounts to assume a feed-forward architecture.  It can be easily shown that $\left({\mathbf \Phi}^{(k)}\right)^{-1}=2 \id{}_{N \times N}- {\mathbf \Phi}^{(k)}$, which readily yields ${\mathbf A}^{(k)}={\mathbf \Phi}^{(k)} {\mathbf \Lambda}^{(k)} \left(2 \id{}_{N \times N}- {\mathbf \Phi}^{(k)} \right)$. The off-diagonal elements of ${\mathbf \Phi}^{(k)}$ define a second set of adjustable parameters to be self-consistently modulated during active training. To implement the learning scheme on these basis, we consider $\vec{x}^{(\ell)}$, the image on the output layer of the input vector $\vec{x}^{(1)}$:

\begin{equation} \label{image}
	\vec{x}^{(\ell)} = f\left(\mathbf{A}^{(\ell-1)}... f\left (\mathbf{A}^{(1)} \vec{x}^{(1)},\beta_1 \right),\beta_{\ell-1} \right)	  
\end{equation}

Since we are dealing with image classification, we can calculate $\vec{z} = softmax(\vec{x}^{(\ell)})$. We will then use $\vec{z}$ to compute the categorical cross-entropy loss function $\text{CCE}(l(\vec{x}^{(1)}), \vec{z})$, where  $l(\vec{x}^{(1)})$ is the label which identifies the category to which $\vec{x}^{(1)}$ belongs, via one-hot encoding \cite{aggarwal2018neural}. 


The loss function can thus be minimized by acting on the spectral parameters, i.e. the ensemble made of non trivial eigenvalues and/or the associated eigendirections.  A straightforward calculation, carried out in the annexed supplementary information, allows one to derive a closed analytical expression for $w^{(k)}_{ij}$, the weights of the edges linking nodes $i$ (belonging to layer $k+1$) and $j$ (sitting on layer $k$) in direct space, as a function of the underlying spectral quantities. In formulae, one gets: 
\begin{equation} \label{w}
	w^{(k)}_{ij} = \left( \lambda^{(k)}_{m(j)}-\lambda^{(k)}_{l(i)} \right)  {\Phi}^{(k)}_{l(i), m(j)}
\end{equation}
where $l(i)=\sum_{s=1}^{k} N_s +i$ and $m(j)=\sum_{s=1}^{k-1} N_s +j$, with $i \in \left(1, ..., N_{k+1} \right)$ and $j \in \left(1, ..., N_k \right)$. In the above expression, $\lambda^{(k)}_{m(j)}$ stand for the first $N_k$  eigenvalues of ${\mathbf \Lambda}^{(k)}$. The remaining $N_{k+1}$ eigenvalues are labelled $\lambda^{(k)}_{l(i)}$.

To help comprehension denote by  $x^{(k)}_j$ the activity on nodes $j$. Then, the activity $x^{(k)}_i$ on node $i$ reads:
\tiny
\begin{equation}
	x^{(k+1)}_i = \sum_{j=1}^{N_k} \left( \lambda^{(k)}_{m(j)}   {\Phi}^{(k)}_{l(i), m(j)} x_j^{(k)} \right) - \lambda^{(k)}_{l(i)} \sum_{j=1}^{N_k} \left( {\Phi}^{(k)}_{l(i), m(j)} x_j^{(k)} \right)
\end{equation} 
\normalsize

The eigenvalues $\lambda^{(k)}_{m(j)}$ modulate the density at the origin, while $\lambda^{(k)}_{l(i)}$ set the excitability of the receiver nodes, weighting the network activity in its immediate neighbourhood. As remarked in \cite{chicchi2021training}, this can be rationalized as the artificial analogue of the {\it homeostatic plasticity}, the strategy used by living neurons to maintain the synaptic basis for learning, respiration, and locomotion \cite{surmeier2004mechanism}. 

Starting from this background, we shall hereafter operate within a simplified setting which is obtained by imposing $\lambda^{(k)}_{m(j)} = 0$. This implies that $\lambda^{(k)}_{l(i)}$ are the sole eigenvalues to be actively involved in the training. As we shall prove, these latter eigenvalues provide an effective criterion to rank a posteriori, i.e. upon training being completed, the relative importance of the nodes belonging to the examined network. Stated differently, nodes can be sorted according to their relevance in carrying out the assigned task. This motivates us to introduce, and thoroughly test, an effective spectral pruning strategy which seeks at removing the nodes deemed unessential, while preserving the overall network classification score. The Methods Section is entirely devoted to explain in detail the proposed strategy, that we shall contextualize with reference to other existing methodologies.   

\section{Conventional Pruning Techniques}
Generally speaking, it is possible to ideally group various approaches for network compression into five different categories: Weights Sharing, Network Pruning, Knowledge Distillation, Matrix Decomposition and Quantization \cite{neill2020overview, cheng2017survey}.

Weights Sharing defines one of the simplest strategies to reduce the number of parameters, while allowing for a robust feature detection. The key idea is to have a shared set of model parameters between layers,  a choice which reflects back in an effective model compression. An immediate example of this methodology are the convolutional neural networks \cite{lecun1989backpropagation}. A refined approach is proposed in Bat et al. \cite{bai2019deep} where a virtual infinitely deep neural network is considered. Further, in Zhang et al. \cite{zhang2018learning} an $\ell_{1}$ group regularizer is exploited to induce sparsity and, simultaneously, identify the subset of weights which can share the same features.

Network Pruning is arguably one of the most common technique to compress Neural Network: in a nutshell it aims at removing a set of weights according to a certain criterion (magnitude, importance, etc). Chang et al. \cite{chang2018prune} proposed an iterative pruning algorithm that exploits a continuously differentiable version of the $\ell_{\frac{1}{2}}$ norm, as a penalty term. Molchanov et al. \cite{molchanov2016pruning} focused on pruning convolutional filters, so as to achieve better inference performances (with a modest impact on the recorded accuracy) in a transfer leaning scenario. Starting from a network fine-tuned on the target task, they proposed an iterative algorithm made up of three main parts: (i) assessing the importance of each convolutional filter on the final performance via a Taylor expansion, (ii) removing the less informative filters and (iii) re-training the remaining filters, on the target task. Inspired by the pioneering work in \cite{frankle2018lottery}, Pau de Jorge et al. \cite{de2020progressive} proved that pruning at initialization leads to a significant performance degradation, after a certain pruning threshold. In order to overcome this limitation they proposed two different methods that enable an initially trimmed weight to be reconsidered during the subsequent training stages.

Knowledge Distillation is yet another technique, firstly proposed by Hinton et al. \cite{hinton2015distilling}. In its simplest version Knowledge Distillation is implemented by combining two objective functions. The first accounts for the discrepancy between the predicted and true labels. The second is the cross-entropy between the output produced by the examined network and that obtained  by running a (generally more powerful) trained model. In \cite{polino2018model} Polino et al. proposed two approaches to mix distillation and quantization (see below): the first method uses the distillation during the training of the so called student network under a fixed quantization scheme while the second  exploits a network (termed the teacher network) to directly optimize the quantization. Mirzadeh et al. \cite{mirzadeh2020improved} analyzed the regime in which knowledge distillation can be properly leveraged. They discovered that the representation power gap of the two networks (teacher and student) should be bounded for the method to yield beneficial effects. To resolve this problem, they inserted an intermediate network (the assistant), which sits in between the teacher and the student, when their associated gap is too large. 

Matrix Decomposition is a technique that remove redundancies in the parameters by the means of a tensor/matrix decomposition. Masana et al. \cite{masana2017domain} proposed a matrix decomposition method for transfer learning scenario. They showed that decomposing a matrix taking into account the activation outperforms the approaches that solely rely on the weights. In \cite{novikov2015tensorizing}, Novikov et al. proposed to replace the dense layer with its Tensor-Train representation \cite{oseledets2011tensor}. Yu et al. \cite{yu2017compressing} introduced a unified framework, integrating the low-rank and sparse decomposition of weight matrices with the feature map reconstructions.

Quantization, as also mentioned above, aims at lowering the number of bits used to represent any given parameter of the network.  Stock et al. \cite{stock2019and} defined an algorithm that quantize the model by minimizing the reconstruction error for inputs sampled from the training set distribution. The same authors also claimed that their proposed method is particularly suited for compressing residual network architectures and that the compressed model proves very efficient when run on CPU. In Banner et al. \cite{banner2018post} a practical 4-bit post-training quantization approach was introduced and tested.
Moreover, a method to reduce network complexity based on node-pruning was presented by He et al. in \cite{He2014}. Once the network has been trained, nodes are classified by means of a node importance function and then removed or retained depending on their score. The authors proposed three different node ranking functions: entropy, output-weights norm (onorm) and input-weights norm (inorm). In particular, the input-weights norm function is defined as the sum of the absolute values of the incoming connections weights. As we will see this latter defines the benchmark model that we shall employ to challenge the performance of the trimming strategy here proposed. Finally, it is worth mentioning the Conditional Computation methods \cite{wang2020deep, Wang_2018_ECCV, bengio2015conditional}: the aim is to dynamically skip part of the network according to the provided input so as to reduce the computational burden.

Summing up, pruning techniques exist which primarily pursue the goal of enforcing a sparsification by cutting links from the trained neural network and have been reviewed above.
In contrast with them, the idea of our method is to a posteriori identify the nodes of the trained network which prove unessential for a proper functioning of the device and cut them out from ensemble made of active units. This yields a more compact neural network, in terms of composing neurons, with unaltered classification performance. The method relies on the spectral learning \cite{spec_learn,chicchi2021training} and exploits the fact that eigenvalues are credible parameters to gauge the importance of a given node among those composing the destination layer. In short, our aim is to make the network more compact by removing nodes classified as unimportant, according to a suitable spectral rating.

\section{Results}
In order to assess the effectiveness of the eigenvalues as a marker of the node's importance (and hence as a potential target for a cogent pruning procedure) we will consider a fully connected feed-forward architecture. Applications of the explored methods will be reported for $\ell=3$ and $\ell>3$ configurations. The nodes that compose the hidden layers are the target of the implemented pruning strategies. As we shall prove, it is possible to get rid of the vast majority of nodes without reflecting in a sensible decrease in the test accuracy, if the filter, either in its pre- or post-training versions, relies on the eigenvalues ranking.

For our test, we used three different datasets of images. The first is the renowned MNIST database of handwritten digits \cite{lecun1998mnist}, the second is Fashion-MNIST (F-MNIST) \cite{xiao2017fashion} (an image dataset of Zalando's items) and the last one is CIFAR-10 \cite{krizhevsky2009learning}. In the main text we report our findings for Fashion-MNIST. Analogous investigations carried out for MNIST and CIFAR10 will be reported as supplementary information. Further, different activation functions have been employed to evaluate the performance of the methods. In the main body of the paper, we will show the results obtained for the ELU. The conclusion obtained when operating with the ReLU and $\tanh$ are discussed in the annexed supplementary material. In the following we will report into two separate sub-sections the results pertaining to either the single or multiple hidden layers settings.

\subsection{Single hidden layer ($\ell=3$)}
\begin{figure}
	\centering
	\includegraphics[width = 0.45\textwidth]{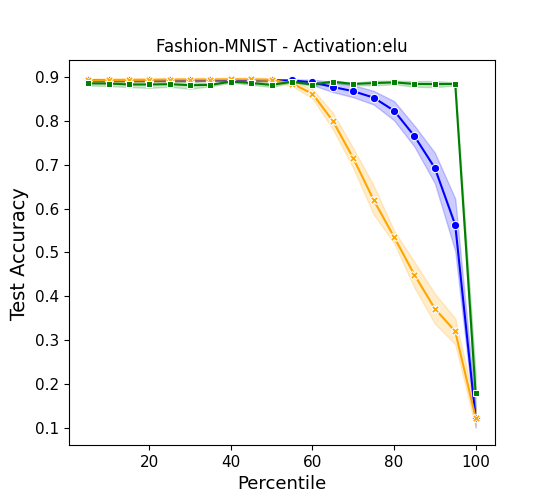}
	\caption{Accuracy on the Fashion-MNIST database with respect to the percentage of trimmed nodes (from the hidden layer), in a three layers feedforward architecture. Here, $N_2=500$, while $N_1=784$ and $N_3=10$, as reflecting the structural characteristics of the data. In orange the results obtained by pruning the network trained in direct space, based on the  absolute value of the incoming connectivity (see main text). In blue, the results obtained when filtering the nodes after a full spectral training (post-training). The curve in green reports the accuracy of the trimmed networks generated upon application of the pre-training filter. Symbols stand for the averaged accuracy computed over 5 independent realizations. The shadowed region is traced after the associated semi-dispersion.}
	\label{f:f-mnist ELU}
\end{figure}

In Figure \ref*{f:f-mnist ELU} the performance of the inspected methods are compared for the minimal case study of a three layers network. The intermediate layer, the sole hidden layer in this configuration, is set to $N_2=500$ neurons. The accuracy of the different methods are compared, upon cutting at different percentile, following the strategies discussed in the Methods. The orange profile is the benchmark model: the neural network is trained in direct space, by adjusting the weights of each individual inter-nodes connection. Then, the absolute value of the incoming connectivity is computed and used as an importance rank of the nodes' influence on the test accuracy. Such a model has been presented and discussed by He et al. in \cite{He2014}. Following this assessment, nodes are progressively removed from the trained network, depending on the imposed percentile, and the ability of the trimmed network to perform the sought classification (with no further training) tested. The same procedure is repeated $5$ times and the mean value of the accuracy plotted. The shaded region stands for the semi dispersion of the measurements. A significant drop of the network performance is found when removing a fraction of nodes larger than 60 \% from the second layer. 

The blue curve Figure \ref*{f:f-mnist ELU} refers instead to the post-processing spectral pruning based on the eigenvalues and identified, as method (ii), in the Methods Section. More precisely, the three layers network is trained by simultaneously acting on the eigenvectors and the eigenvalues of the associated transfer operators, as illustrated above. The accuracy displayed by the network trained according to this procedure is virtually identical to that reported when the learning is carried out in direct space, as one can clearly appreciate by eye inspection of Figure \ref*{f:f-mnist ELU}. Removing the nodes based on the magnitude their associated eigenvalues, allows one to keep stable (practically unchanged) classification performance for an intermediate layer that is compressed of about 70\% of its original size. In this case the spectral pruning is operated as a post-processing filter, meaning that the neural network is only trained once, before the nodes' removal takes eventually place.

At variance,  the green curve in Figure \ref*{f:f-mnist ELU} is obtained following method (i) from the Methods Section, which can be conceptualized as a pre-training manipulation. Based on this strategy, we first train the network on the set of tunable eigenvalues, than reduce its size by performing a compression that reflects the ranking of the optimized eigenvalues and then train again the obtained network by acting uniquely on the ensemble of residual eigenvectors. The results reported in Figure \ref*{f:f-mnist ELU} indicate that, following this procedure, it is indeed possible to attain astoundingly compact networks with unaltered classification abilities. Moreover, the total number of parameters that need to be tuned following this latter procedure is considerably smaller than that on which the other methods rely. This is due to the fact that only the random directions (the eigenvectors) that prove relevant for discrimination purposes (as signaled by the magnitude of their associated eigenvalues) undergoes the second step of the optimization. This method can also be seen as a similar kind of \cite{frankle2018lottery}. As a matter of fact, the initial training of the eigenvalues uncovers a sub-network that, once trained, obtains performances comparable to the original model. More specifically, the uncovered network can be seen as a \textit{winning ticket} \cite{frankle2018lottery}. That is, a sub-network with an initialization particularly suitable for carrying out a successful training.

Next, we shall generalize the analysis to the a multi-layer setting ($\ell>3$), reaching analogous conclusions. 

\subsection{Multiple hidden layers ($\ell>3$)}
Quite remarkably, the results achieved in the simplified context of a single hidden layer network 
also apply within the framework of a multi-layers setting.\\
To prove this statement we set to consider a $\ell=5$ feedforward neural network with ELU activation. Here, $N_1=784$ and $N_5=10$ as reflecting the specificity of the employed dataset. 
The performed tests follows closely those reported above, with the notable difference that now the ranking of the eigenvalues is operated on the pool of $ N_2 + N_3 + N_4 $ neurons that compose the hidden bulk of the trained network. In other words, the selection of the neuron to be removed is operated after a global assessment, i.e. scanning across the full set of nodes, without any specific reference to an a priori chosen layer.  

\begin{figure}[!ht]
	\centering
	\includegraphics[width = 0.45\textwidth]{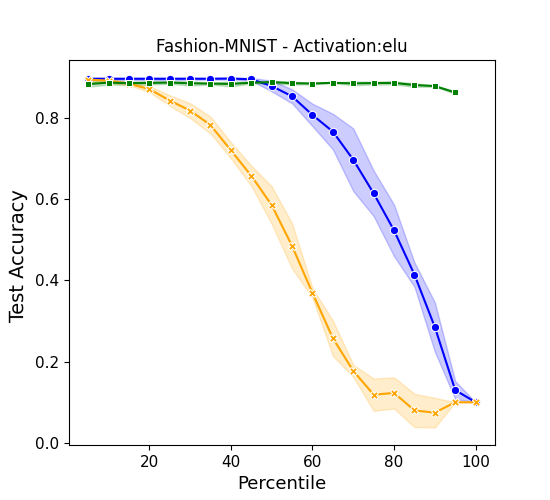}
	\caption{Accuracy on the Fashion-MNIST database with respect to the percentage of pruned nodes (from the hidden layers), in a five layers feedforward architecture. Here, $N_2=N_3=N_4=500$, while $N_1=784$ and $N_5=10$, as reflecting the structural characteristics of the data. Symbols and colors are chosen as in Figure \ref*{f:f-mnist ELU}.}
	\label{f:multi f-mnist ELU}
\end{figure}

In Figure \ref{f:multi f-mnist ELU} the results of the analysis are reported, assuming $N_2=N_3=N_4=500$. The conclusions are perfectly in line with those reported above for the one layer setting, except for the fact that now the 
improvement of the spectral pruning over the benchmark reference are even superior. The orange curve drops at percentile 20, while the blue begins its descent at about 60 \%. The green curve, relative to the sequential two steps training, stays stably horizontal up to about 90 \%. 

\section{Conclusions}
In this paper we have discussed a relevant byproduct of a spectral approach to the learning of deep neural networks. The eigenvalues of the transfer operator that connects adjacent stacks in a multi-layered architecture provide an effective measure of the nodes importance in handling the information processing. By exploiting this fact we have introduced and successfully tested two distinct procedures to yield compact networks --in terms of number of computing neurons-- which perform equally well than their untrimmed original homologous. 
One procedure (referred as (ii) in the description) is acknowledged as a post processing method, in that it acts on a multi-layered network downstream of training. The other (referred as (i)) is based on a sequence of two nested operations. First the eigenvalues are solely trained. After the spectral pruning took place, a second step in the optimization path seeks to adjust the entries of the eigenvectors that populate a trimmed space of reduced dimensionality. The total number of trained parameters is small as compared to that involved when the pruning acts as a post processing filter. Despite that, the two steps pre-processing protocol yields compact devices which outperform those obtained with a single post-processing removal of the unessential nodes. 

As a benchmark model, and for a neural network trained in direct space, we decided to rank the nodes importance based on the absolute value of the incoming connectivity. This latter appeared as the obvious choice, when aiming at gauging the local information flow in the space of the nodes, see also \cite{He2014}. In principle, one could consider to diagonalizing the transfer operators as obtained after a standard approach to the training and make use of the computed eigenvalues to a posteriori sort the nodes relevance. This is however not possible as the transfer operator that links a generic layer $k$ to its adjacent counterpart $k+1$, as follows the training performed in direct space, is populated only below the diagonal, with all diagonal entries identically equal zero. All associated eigenvalues are hence are zero and they provide no information on the relative importance of the nodes of layer $k+1$, at variance with what happens when the learning is carried out in the reciprocal domain.

Summing up, by reformulating the training of neural networks in spectral space, we identified a set of sensible scalars, the eigenvalues of suitable operators, that unequivocally correlate with the influence of the nodes within the collection. This observation translates in straightforward procedures to generate efficient networks that exploit a reduced number of computing units. Tests performed on different settings corroborate this conclusions. As an interesting extension, we will show in the supplementary information that a suitable regularization of the eigenvalues yields a general improvement of the proposed method.

\section{Methods}\label{method}
We detail here the spectral procedure to make a trained network smaller, while preserving its ability to perform classification. 

To introduce the main idea of the proposed method, we make reference to formula (\ref{w}) and assume the setting where $\lambda^{(k)}_{m(j)}=0$. The information travelling from layer $k$ to layer $k+1$ gets hence processed as follows: first, the activity on  the departure node $j$ is modulated by a multiplicative scaling factor ${\Phi}^{(k)}_{l(i), m(j)}$, specifically linked to the selected $(i,j)$ pair. Then, all incoming (and rescaled) activities reaching the destination node $i$ are summed together and further weighted via the scalar quantity $\lambda^{(k)}_{l(i)}$. This latter eigenvalue, downstream of the training, can be hence conceived as a distinguishing feature of node $i$ of layer $k+1$. Assume for the moment that ${\Phi}^{(k)}_{l(i), m(j)}$ are drawn from a given distribution and stay put during optimization. Then, every individual neuron bound to layer $k+1$ is statistically equivalent (in terms of incoming weights) to all other  nodes, belonging to the very same layer. The eigenvalues 
$\lambda^{(k)}_{l(i)}$  gauge therefore the relative importance of the nodes, within a given stack, and as reflecting the (randomly generated) web of local  inter-layer connections (though statistically comparable).  Large values of $|\lambda^{(k)}_{l(i)}|$ suggest that node $i$ on layer $k+1$ plays a central role in the economy of the neural network functioning. This is opposed to the setting when $|\lambda^{(k)}_{l(i)}|$ is found to be small. Stated differently, the subset of trained eigenvalues provide a viable tool to rank the nodes according to their degree of importance. As such, they can be used as reference labels to make decision on the nodes that should be retained in  a compressed analogue of the trained neural network, with unaltered classification performance. As empirically shown in the Results section with reference to a variegated set of  applications, the sorting of the nodes based on the optimized eigenvalues turns out effective also when the eigenvectors get simultaneously trained, thus breaking, at least in principle, statistical invariance across nodes. 

As we will clarify, the latter setting translates in a post-training spectral pruning strategy, whereas the former materializes in a rather efficient pre-training procedure. The non linear activation function as employed in the training scheme leaves a non trivial imprint, which has to be critically assessed. 

More specifically, in carrying out the numerical experiments here reported  we considered two distinct settings, as listed below:

\begin{itemize}
	\item{(i)} As a first step, we will begin by considering a deep neural network made of $N$ neurons organized in $\ell$ layers. The network will be initially  trained by  solely leveraging on the set of tunable eigenvalues. Then, we will proceed by progressively removing the neurons depending on their associated eigenvalues (as in the spirit discussed above). The trimmed network, composed by a total of $M<N$ units, still distributed in $\ell$ distinct layers, can be again trained acting now on the eigenvectors, while keeping the eigenvalues frozen to the earlier determined values. This combination of steps, which we categorize as pre-training, yields a rather compact neural network ($M$ can be very small) which performs equally well than its fully trained analogue made of $N$ computing nodes.
	
	\item{(ii)} We begin by constructing a deep neural network made of $N$ neurons organized in $\ell$ layers. This latter undergoes a full spectral training, which optimizes simultaneously eigenvectors and the eigenvalues. The trained network can be compressed, by pruning the nodes which are associated to eigenvalues (see above) with magnitude smaller that a given threshold. This is indeed a post-training pruning strategy, as it acts {\it ex post} on a fully trained device.
\end{itemize}

\begin{figure}
		\centering
		\includegraphics[width = 0.5\textwidth]{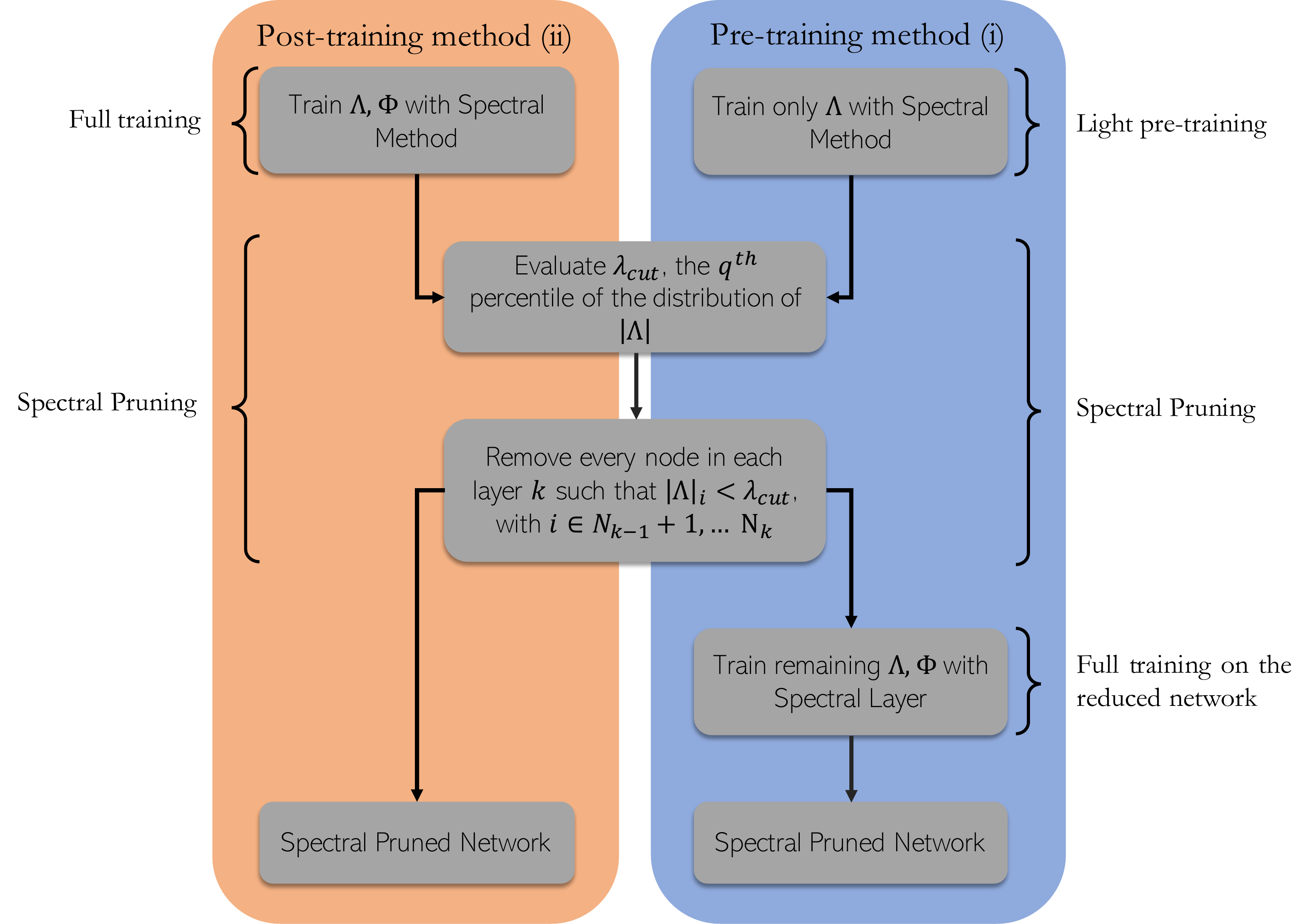}
		\caption{Flowchart of the pre- and post- training pruning strategies as presented in section \ref{method}. }
		\label{f:flowchart}
\end{figure}

To evaluate the performance of the proposed spectral pruning strategies (schematically represented in the flowchart of Figure \ref{f:flowchart}), we also introduced a reference benchmark model. This latter can be conceptualized as an immediate overturning of the methods in direct space. Simply stated, we train the neural network in the space of the nodes, by using standard approaches to the learning. Then, we classify the nodes in terms of their relevance using a proper metric to which shall make reference below, and consequently trim the nodes identified as less important. When adopting the spectral viewpoint, one can rely on the eigenvalues to rank the nodes importance. As remarked above, in fact, the eigenvalues at the receiver nodes set a local scale for the incoming activity, the larger the eigenvalue (in terms of magnitude) the more important the role played by the processing unit. As a surrogate of the eigenvalues, when anchoring the train in direct space, we can consider the quantity $\sum_{j=1}^{N_k} |w_{ij}|$, for each neuron $i$ belonging to layer $k+1$, see also \cite{He2014}. The absolute value prevents mutual cancellations of sensible contributions bearing opposite signs, which could incidentally hide the actual importance of the examined node. 

In all explored cases, the pruning is realized by imposing a threshold on the reference indicator (be it the magnitude of the eigenvalues or the cumulated flux of incoming --and made positive-- weights). Pointedly,  the respective indicator is extracted for every node in the arrival layer. Then a percentile $q$ is chosen and the threshold fixed to the $q$-th percentile. Nodes displaying an indicator below the chosen threshold are removed and the accuracy of the obtained (trimmed) neural network assessed on the test-set. The codes employed, as well as a notebook to reproduce our results, can be found in the public repository of this project \footnote{\url{https://github.com/Buffoni/spectral_learning}}.

\bibliographystyle{unsrt}
\bibliography{references}

\section*{Authors Contributions}
LB and EC conceived the idea of the work. All authors participated in writing the code.
EC and LG performed the experiments on the various datasets.
DF supervised the project.
All authors contributed to the writing of the manuscript.

\section*{Competing Interests}
The authors declare no competing interests.

\appendix

\section{Analytical characterisation of inter-nodes weights in direct space}
In the following, we will derive Eq. 2 in the main body of the paper. We begin by recalling that ${\mathbf A}^{(k)}$ is a $N \times N$ matrix. From ${\mathbf A}^{(k)}$ we select a   square sub-block of size $(N_k+N_{k+1}) \times (N_k+N_{k+1})$, formed by the elements ${\mathbf A}^{(k)}_{i',j'}$ with $i'=\sum_{s=1}^{k-1}N_s+i$ and $j'=\sum_{s=1}^{k-1}N_s+j$, with $i=1,..., N_k+N_{k+1}$, $j=1,..., N_k+N_{k+1}$. We use ${\mathbf A}^{(k)}$ to identify the obtained matrix and proceed in analogy for ${\mathbf \Lambda}^{(k)}$ and ${\mathbf \Phi}^{(k)}$. Then:
\begin{equation} \label{decomp}
	\begin{aligned}
		A_{i j}^{(k)}=& \left[{\mathbf \Phi}^{(k)} {\mathbf \Lambda}^{(k)}\left(2 I-{\mathbf \Phi}^{(k)}\right)\right]_{i j}\\
		=& \left[2 {\mathbf \Phi}^{(k)} {\mathbf \Lambda}^{(k)}\right]_{i j}-\left[{\mathbf \Phi}^{(k)} {\mathbf \Lambda}^{(k)} {\mathbf \Phi}^{(k)}\right]_{i j}\\
		=& \alpha_{i j}^{(k)}-\beta_{i j}^{(k)}
	\end{aligned}
\end{equation}                                        

From hereon, we will omit the apex $(k)$.  Assume $ \lambda_1 \dots \lambda_{N_k+N_{k+1}} $ to identify the eigenvalues of the transfer operator ${\mathbf A}$, namely the diagonal entries of $\Lambda $. Hence, $ \Lambda _{ij} = \sum_{j = 1}^{N_k+N_{k+1}} \delta_{ij}\lambda_j $.\\
The quantities $ \alpha_{ij} $ and $ \beta_{ij} $ read:
\begin{equation*}
	\begin{aligned}
		\alpha_{ij} &=2 \sum_{k=1}^{N_k+N_{k+1}} \Phi_{i k} \lambda_{k} \delta_{k j}=2 \Phi_{i j} \lambda_{j} \\
		\beta_{ij} &=\sum_{k,m = 1}^{N_k+N_{k+1}} \Phi_{i k} \lambda_{k} \delta_{k m} \Phi_{m j} \\
		&=\sum_{m \in \ii \cup \jj} \delta_{i m} \lambda_{m} \Phi_{m j}
	\end{aligned} 
\end{equation*}
where $j \in \mathcal{J}=\left(1, ..., N_k \right)$ refer to the nodes at the departure layer ($k$), whereas $i \in \mathcal{I}=\left(N_k+1,..., N_k+N_{k+1} \right)$ stand for those at arrival. Hence, $\ii \cup \jj = [1, ..., N_k+N_{k+1}]$. The above expression for $\beta_{ij}$ can be further manipulated to eventually yield
\begin{equation*}
	\begin{aligned}
		\beta_{i j} &= \sum_{m \in J} \Phi_{i m} \lambda_{m} \Phi_{m j}+\sum_{m \in I} \Phi_{i m} \lambda_{m} \Phi_{m j}\\
		&= \Phi_{i j} \lambda_{j}+\lambda_{i} \Phi_{i j}
	\end{aligned}
\end{equation*}
and therefore:
\eqref{decomp} as
\begin{equation}\label{pesi}
	\begin{aligned}
		\alpha_{i j}-\beta_{i j} &= 2 \Phi_{i j} \lambda_{j}-\Phi_{i j} \lambda_{j}-\lambda_{i} \Phi_{i j} \\
		&= (\lambda_j-\lambda_i) \phi_{ij}
	\end{aligned}
\end{equation}

From the above expression, one obtains the sought equation, after redefining the index $i$ to have it confined in the interval  $[1, ..., N_{k+1}]$. By definition, the matrix of the weights, $\mathbf{w}$, is in fact a $N_k \times N_{k+1}$ matrix.

\section{MNIST and Fashion-MNIST: single hidden layer with different activation functions.}
\begin{figure*}[!ht]
	\centering
	\begin{subfigure}[H]{0.64\columnwidth}
		\centering
		\includegraphics[width=1.15\linewidth]{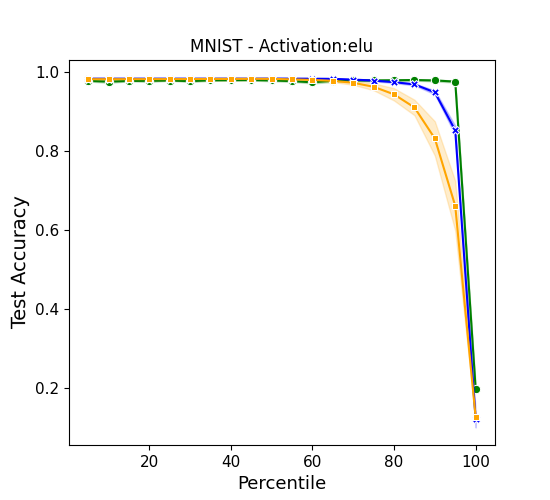}
		\caption{}
		\label{fig:MNISTelu}
	\end{subfigure}
	\hfill
	\begin{subfigure}[H]{0.64\columnwidth}
		\centering
		\includegraphics[width=1.15\linewidth]{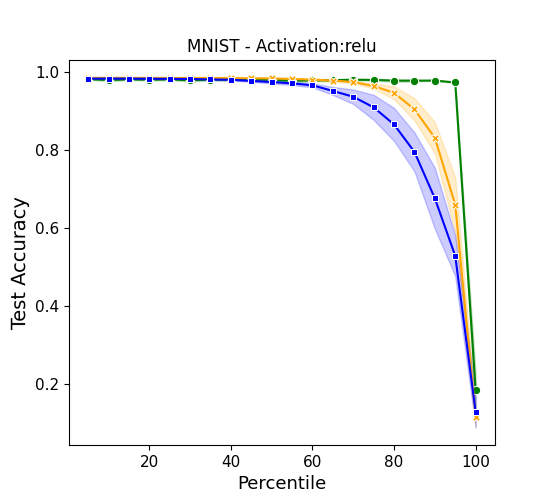}
		\caption{}
		\label{fig:MNISTrelu}
	\end{subfigure}
	\hfill
	\begin{subfigure}[H]{0.64\columnwidth}
		\centering
		\includegraphics[width=1.15\linewidth]{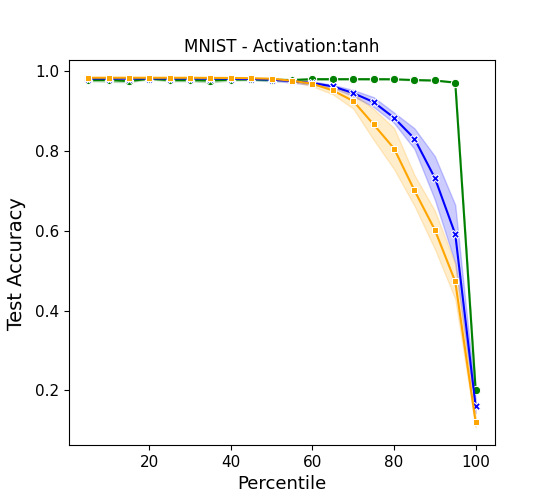}
		\caption{}
		\label{fig:MNISTtanh}
	\end{subfigure}
	\caption{Accuracy on the MNIST database with respect to the percentage of trimmed nodes (selected from the 500 neurons that compose the sole hidden layer), in a three layers feedforward architecture. The results reported in each panel refer to a different selection of the nonlinear activation functions, respectively ELU (a), ReLU (b) and $\tanh$ (c). In orange, the results obtained by using the trimming procedure based on the absolute value of the incoming connectivity. In blue, the results obtained when filtering the nodes after a full spectral training (post-training). The curve in green displays the accuracy of the trimmed networks generated upon application of the pre-training filter. In this case, the examined network is initially trained on the set of eigenvalues, while keeping the eigenvectors frozen. After having removed unessential nodes, based on their associated eigenvalues, the network undergoes another training phase that is solely targeted to adjusting the entries of the residual eigenvectors. The shadowed region represents the semi-dispersion over 5 independent realizations. When using the Relu function, trimming on the absolute value of the incoming connectivity yields slightly better results than what found when using the post-training spectral filter. The two stages spectral trimming proves always more effective.}
\end{figure*}

\begin{figure*}[!ht]
	\centering
	\begin{subfigure}[H]{0.75\columnwidth}
		\includegraphics[width=1.15\linewidth]{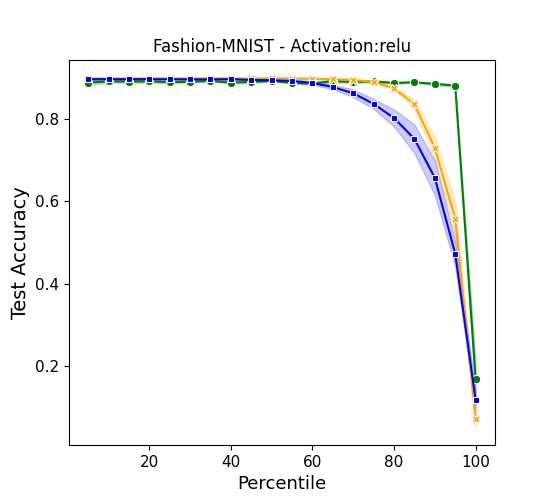}
		\caption{}
		\label{fig:FMNISTrelu}
	\end{subfigure}
	\hfill
	\begin{subfigure}[H]{0.75\columnwidth}
		\includegraphics[width=1.15\linewidth]{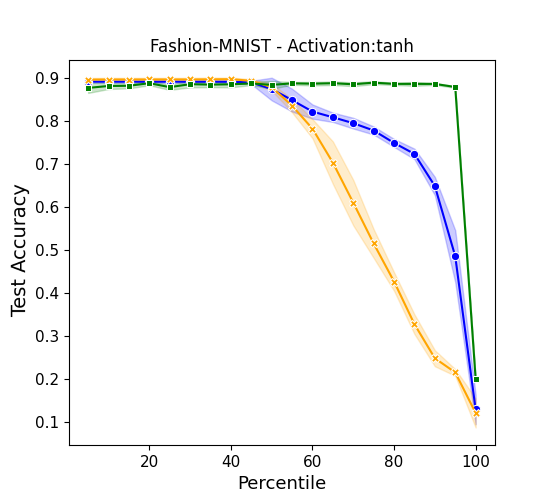}
		\caption{}
		\label{fig:FMNISTtanh}
	\end{subfigure}
	\caption{Accuracy on the Fashion-MNIST database with respect to the percentage of trimmed nodes (selected from the 500 neurons that compose the sole hidden layer), in a three layers feedforward architecture. The results reported in each panel refer to a different selection of the nonlinear activation functions, respectively ReLU (b) and $\tanh$ (c). Symbols and conclusions are in line with those reported for the case of MNIST.}
\end{figure*}

We shall here report (see Figures \ref{fig:MNISTelu}, \ref{fig:MNISTrelu}, \ref{fig:MNISTtanh}, \ref{fig:FMNISTrelu} and \ref{fig:FMNISTtanh}) on the performance of the proposed trimming strategies, as applied to MNIST and Fashion-MNIST, for a single hidden layer architecture and beyond the setting reported in the main body of the paper. In particular, we will assume (i) ELU, $\tanh$ and ReLU for MNIST (ii) $\tanh$ and ReLU activation function for Fashion-MNIST (the ELU activation was employed in the main text). Here, $N_2=500$, while $N_1=784$ and $N_3=10$.

\section{MNIST and Fashion-MNIST: multiple hidden layers with different activation functions.}
\begin{figure*}[!ht]
	\centering
	\begin{subfigure}[H]{0.64\columnwidth}
		\centering
		\includegraphics[width=1.15\linewidth]{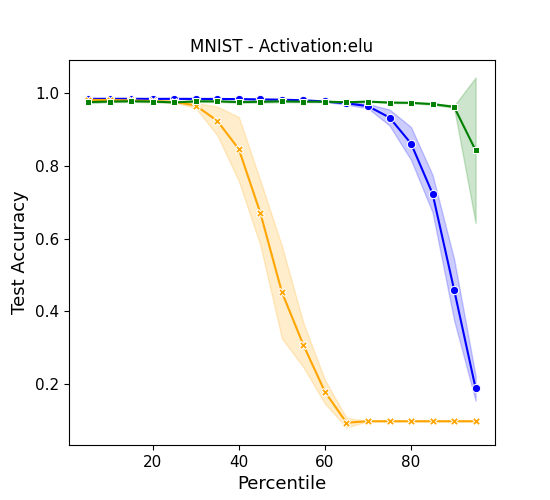}
		\caption{}
		\label{fig:MNISTelu-multi}
	\end{subfigure}
	\hfill
	\begin{subfigure}[H]{0.64\columnwidth}
		\centering
		\includegraphics[width=1.15\linewidth]{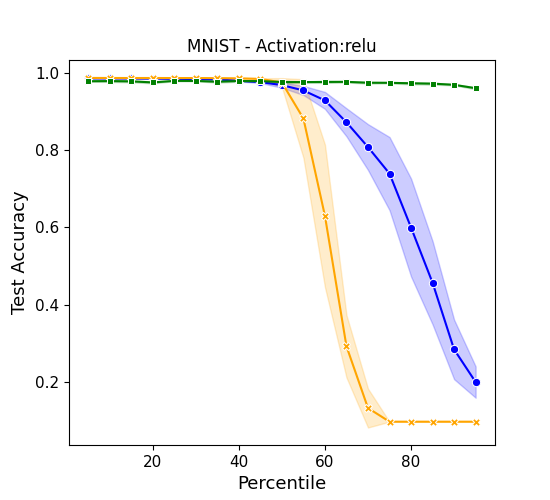}
		\caption{}
		\label{fig:MNISTrelu-multi}
	\end{subfigure}
	\hfill
	\begin{subfigure}[H]{0.64\columnwidth}
		\centering
		\includegraphics[width=1.15\linewidth]{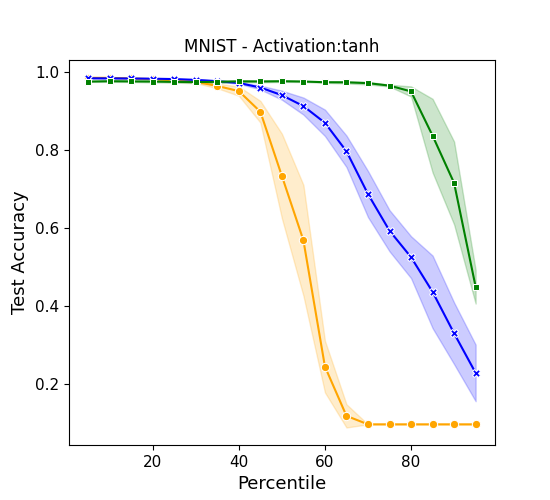}
		\caption{}
		\label{fig:MNISTtanh-multi}
	\end{subfigure}
	\caption{Accuracy on the MNIST database with respect to the percentage of trimmed nodes (from the set of $N_2+N_3+N_4$ neurons). The results in each panel refer to different choices of the non linear function, ELU (a), ReLU (b) and $\tanh$ (c). Symbols are chosen as for the case of the single hidden layer setting. It should be remarked that the spectral trimming strategies proves definitely more effective than the benchmark model anchored to direct space, also when the Relu function is employed, in the case of multiple hidden layers.}
\end{figure*}

\begin{figure*}[!ht]
	\centering
	\begin{subfigure}[H]{0.75\columnwidth}
		\includegraphics[width=1.15\linewidth]{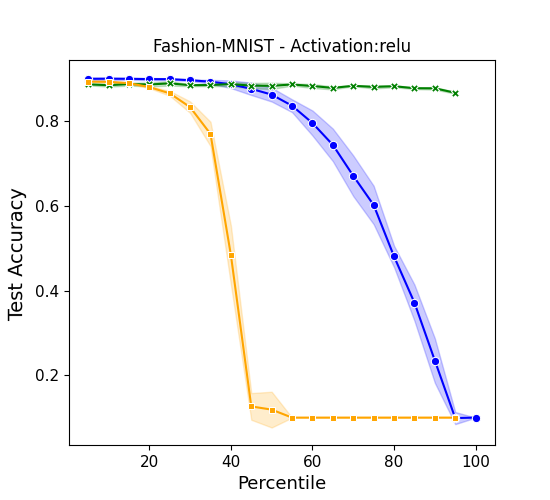}
		\caption{}
		\label{fig:FMNISTrelu-multi}
	\end{subfigure}
	\hfill
	\begin{subfigure}[H]{0.75\columnwidth}
		\includegraphics[width=1.15\linewidth]{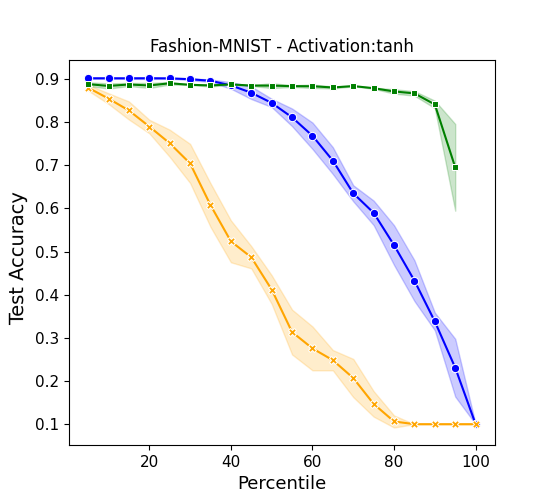}
		\caption{}
		\label{fig:FMNISTtanh-multi}
	\end{subfigure}
	\caption{Accuracy on the Fashion-MNIST database with respect to the percentage of trimmed nodes (from the set of $N_2+N_3+N_4$ neurons). The results in each panel refer to different choices of the non linear activation function, ReLU (a) and $\tanh$ (b). For the symbols, see the caption of the Figures above. Also in this case the spectral filters prove always superior.}
\end{figure*}

We will here generalize the analysis carried out in the preceding section to the case of a multilayered ($\ell>3$) architecture (see Figures \ref{fig:MNISTelu-multi}, \ref{fig:MNISTrelu-multi}, \ref{fig:MNISTtanh-multi}, \ref{fig:FMNISTrelu-multi} and \ref{fig:FMNISTtanh-multi}). In line with the choice operated in the main body of the paper, we will assume a five layered deep neural network with  $N_2=N_3=N_4=500$, and $N_1=784$ and $N_5=10$.

\section{Testing the trimming strategies on CIFAR10 dataset.}
\begin{figure*}[!ht]
	\centering
	\begin{subfigure}[H]{0.64\columnwidth}
		\includegraphics[width=1.15\linewidth]{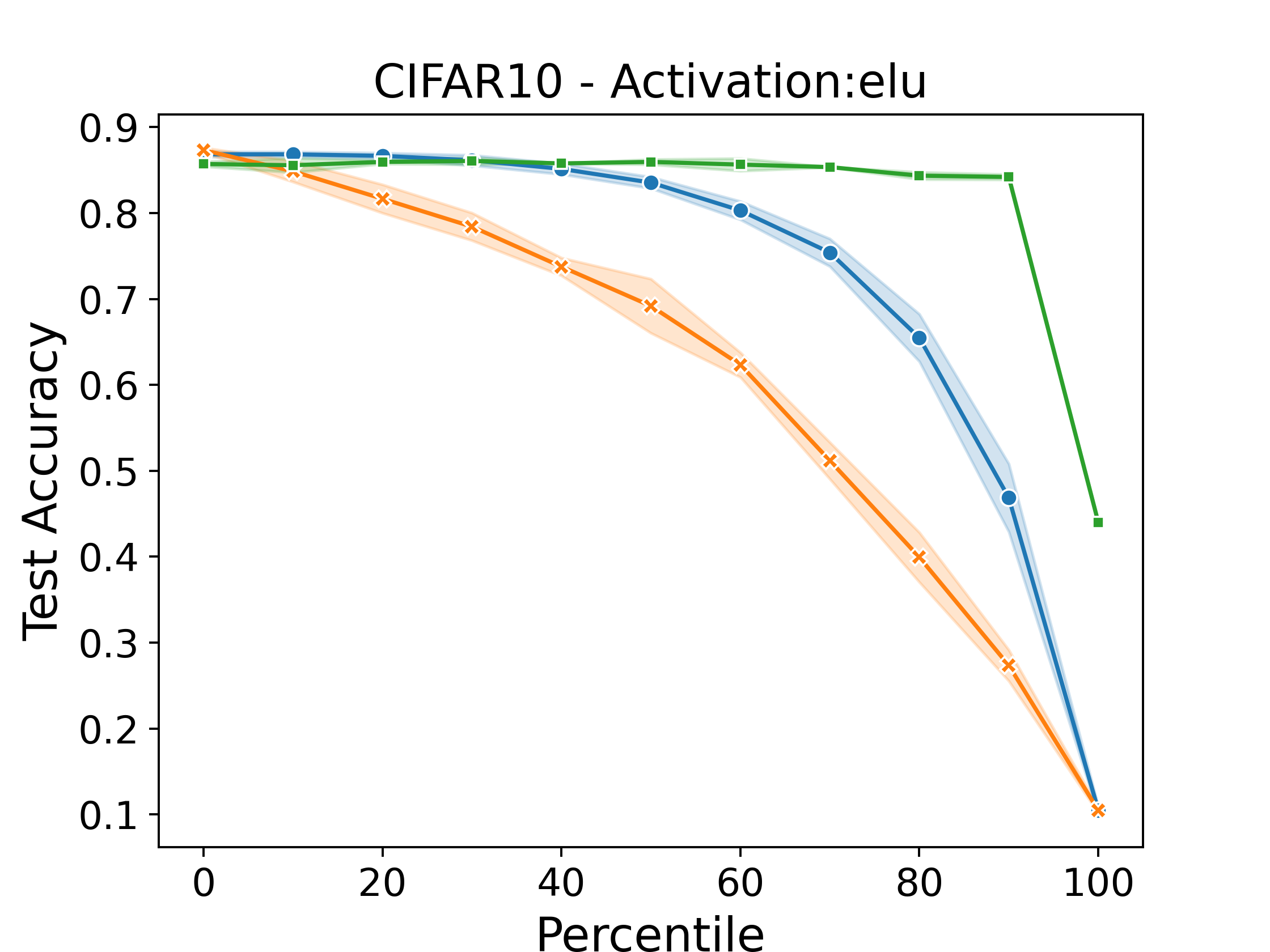}
		\caption{}
		\label{fig:CIFARelu-multi}
	\end{subfigure}
	\hfill
	\begin{subfigure}[H]{0.64\columnwidth}
		\includegraphics[width=1.15\linewidth]{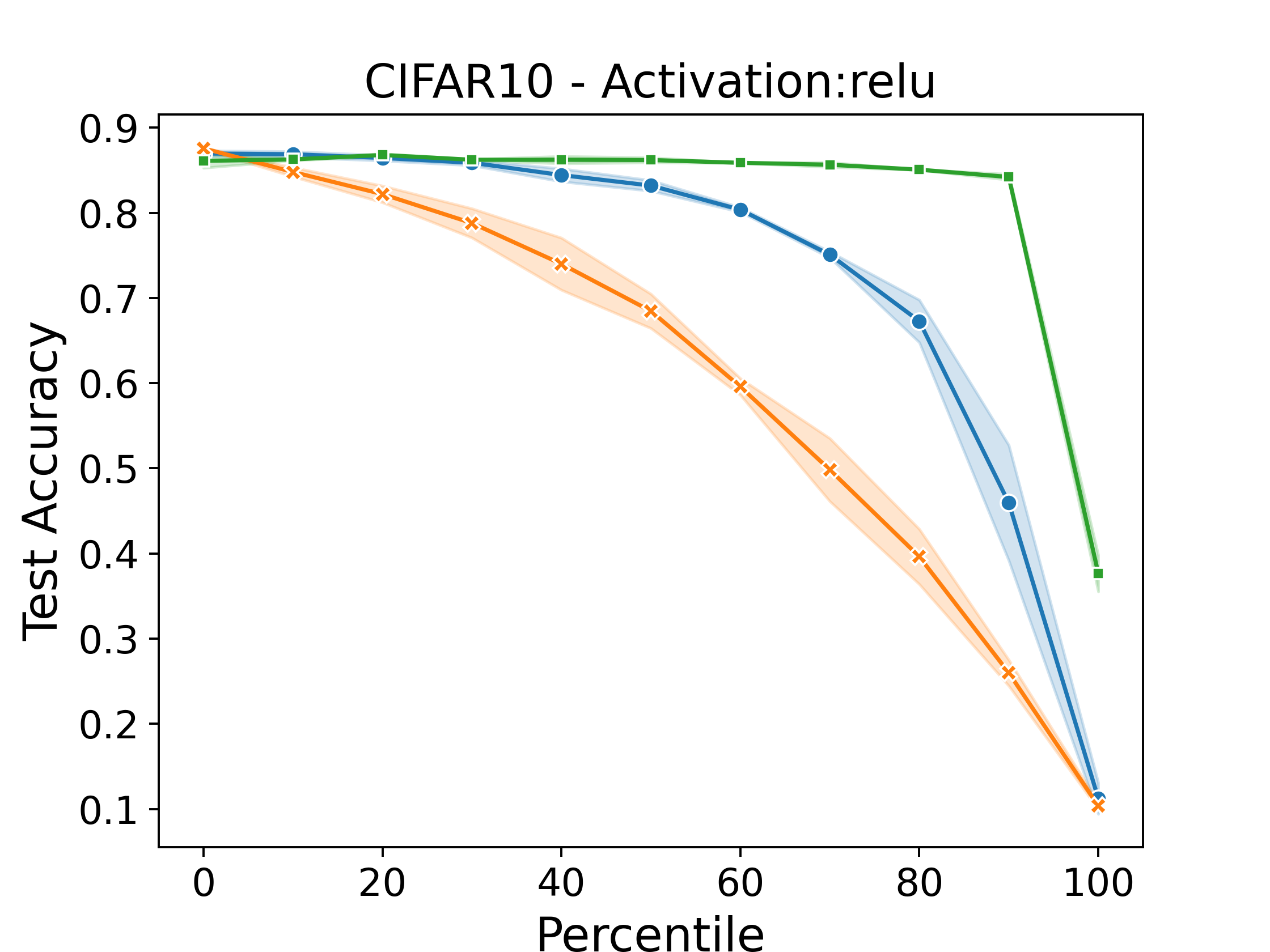}
		\caption{}
		\label{fig:CIFARrelu-multi}
	\end{subfigure}
	\hfill
	\begin{subfigure}[H]{0.64\columnwidth}
		\includegraphics[width=1.15\linewidth]{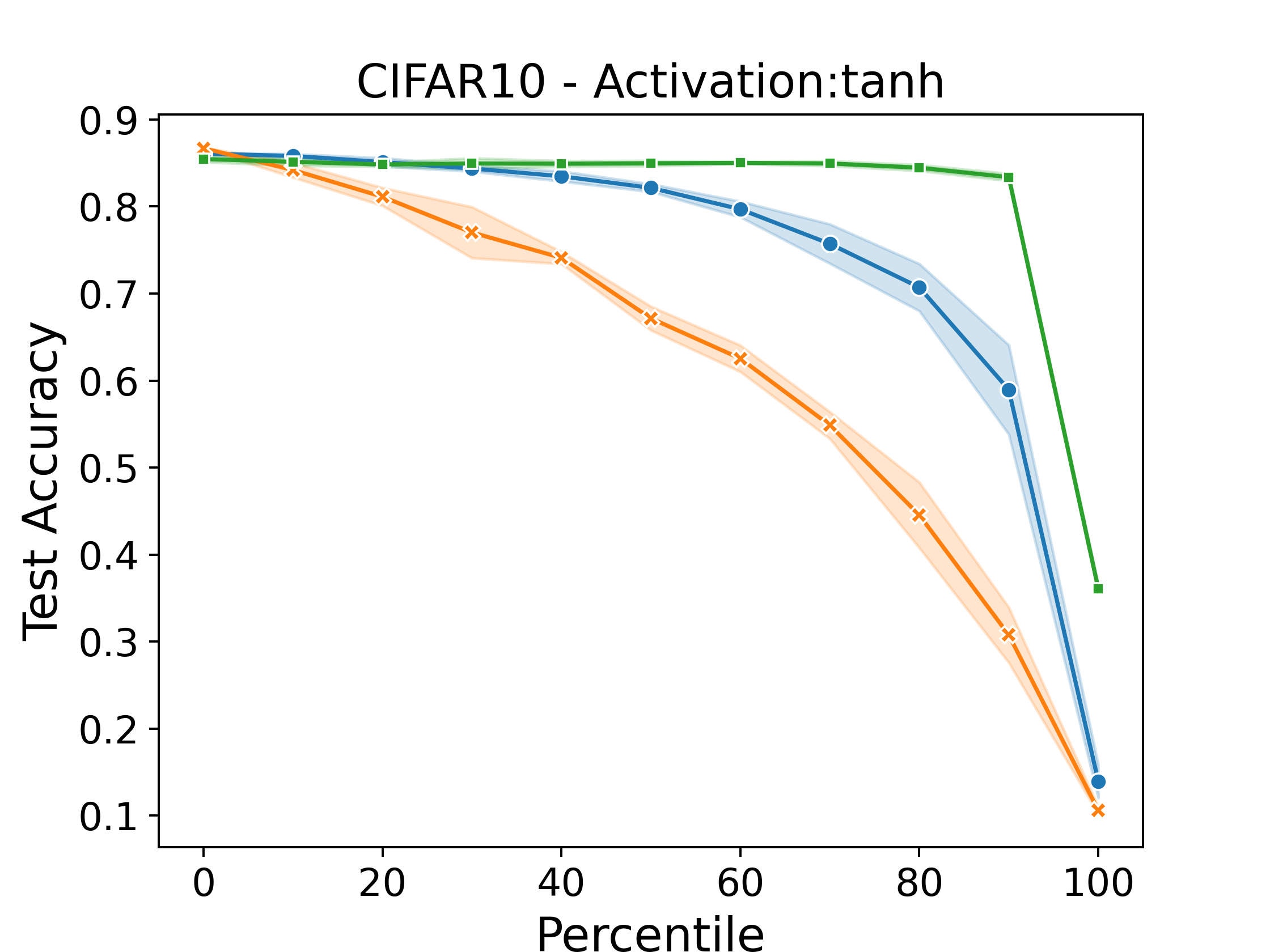}
		\caption{}
		\label{fig:CIFARtanh-multi}
	\end{subfigure}
	\caption{Accuracy on the CIFAR10 database with respect to the percentage of trimmed nodes (from the $ \ell-1$ layer). The results in each panel refer to different non linear functions, respectively ELU (a), ReLU (b) and $\tanh$ (c). Symbols are chosen in analogy with the above (the result drawn in green are based on two different runs).}
\end{figure*}

\begin{figure*}[!ht]
	\centering
	\begin{subfigure}{0.64\columnwidth}
		\centering
		\includegraphics[width=1.17\columnwidth]{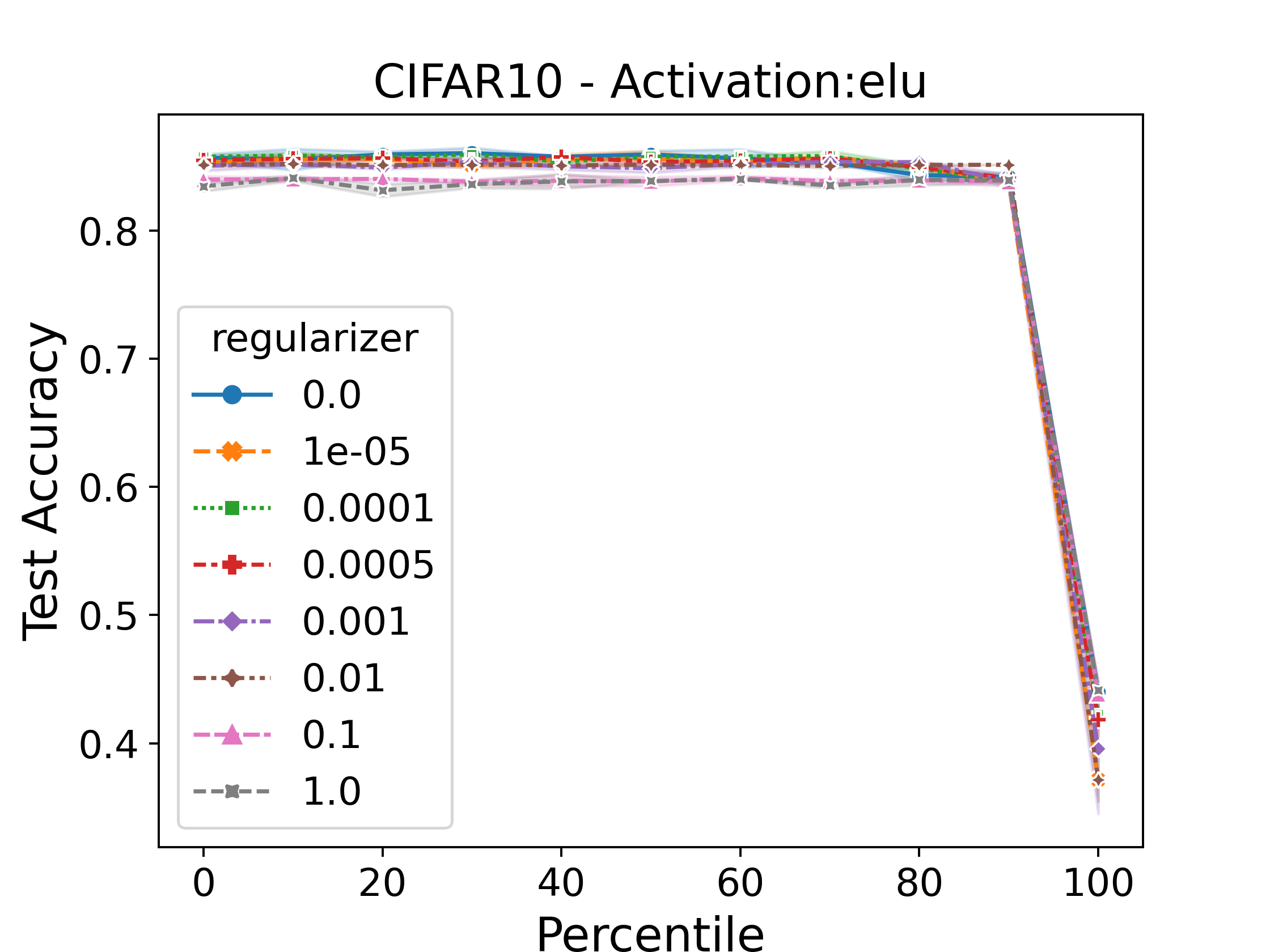}
		\caption{}
	\end{subfigure}
	\hfill
	\begin{subfigure}{0.64\columnwidth}
		\centering
		\includegraphics[width=1.17\columnwidth]{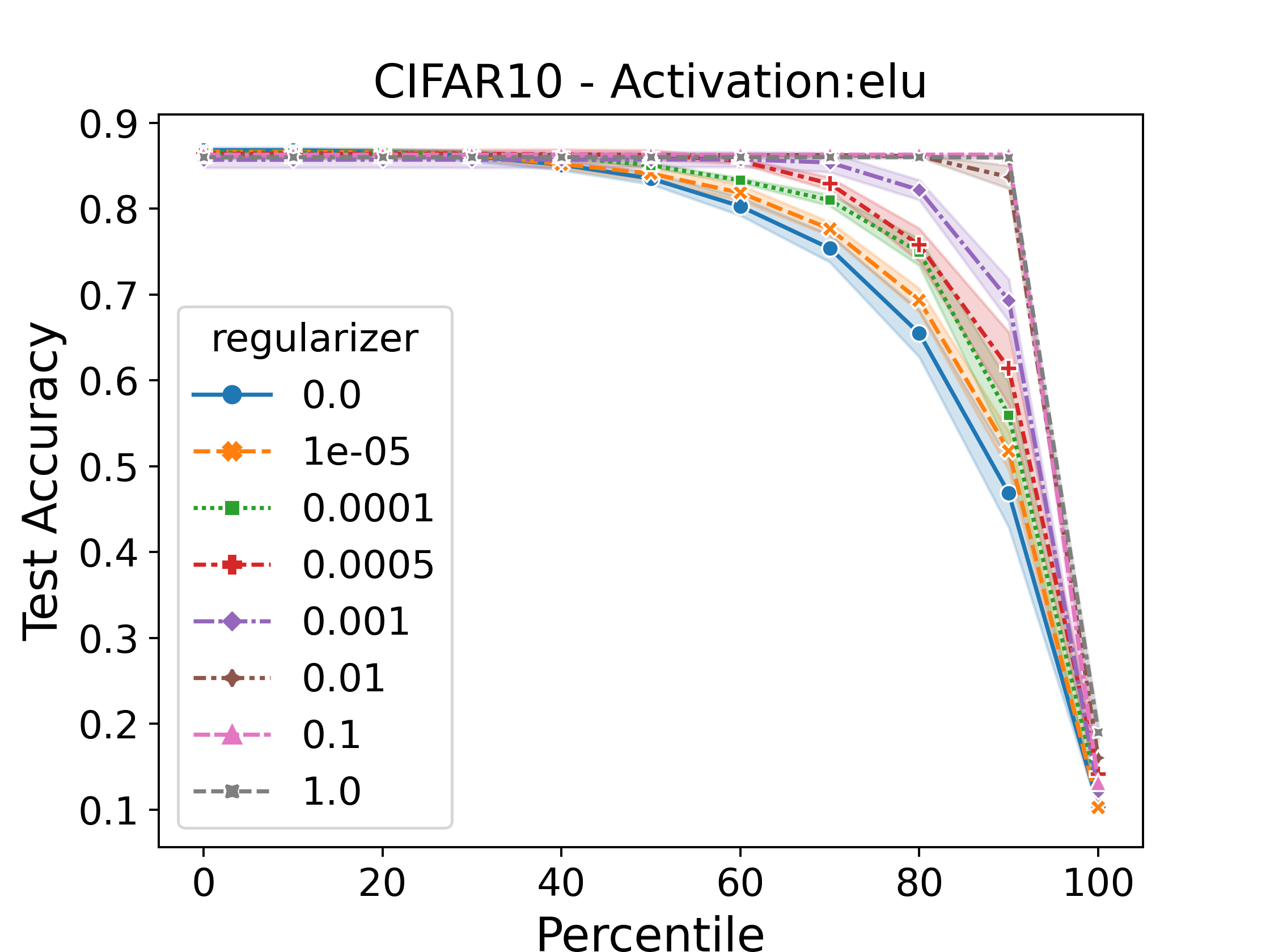}
		\caption{}
	\end{subfigure}
	\hfill
	\begin{subfigure}{0.64\columnwidth}
		\centering
		\includegraphics[width=1.17\columnwidth]{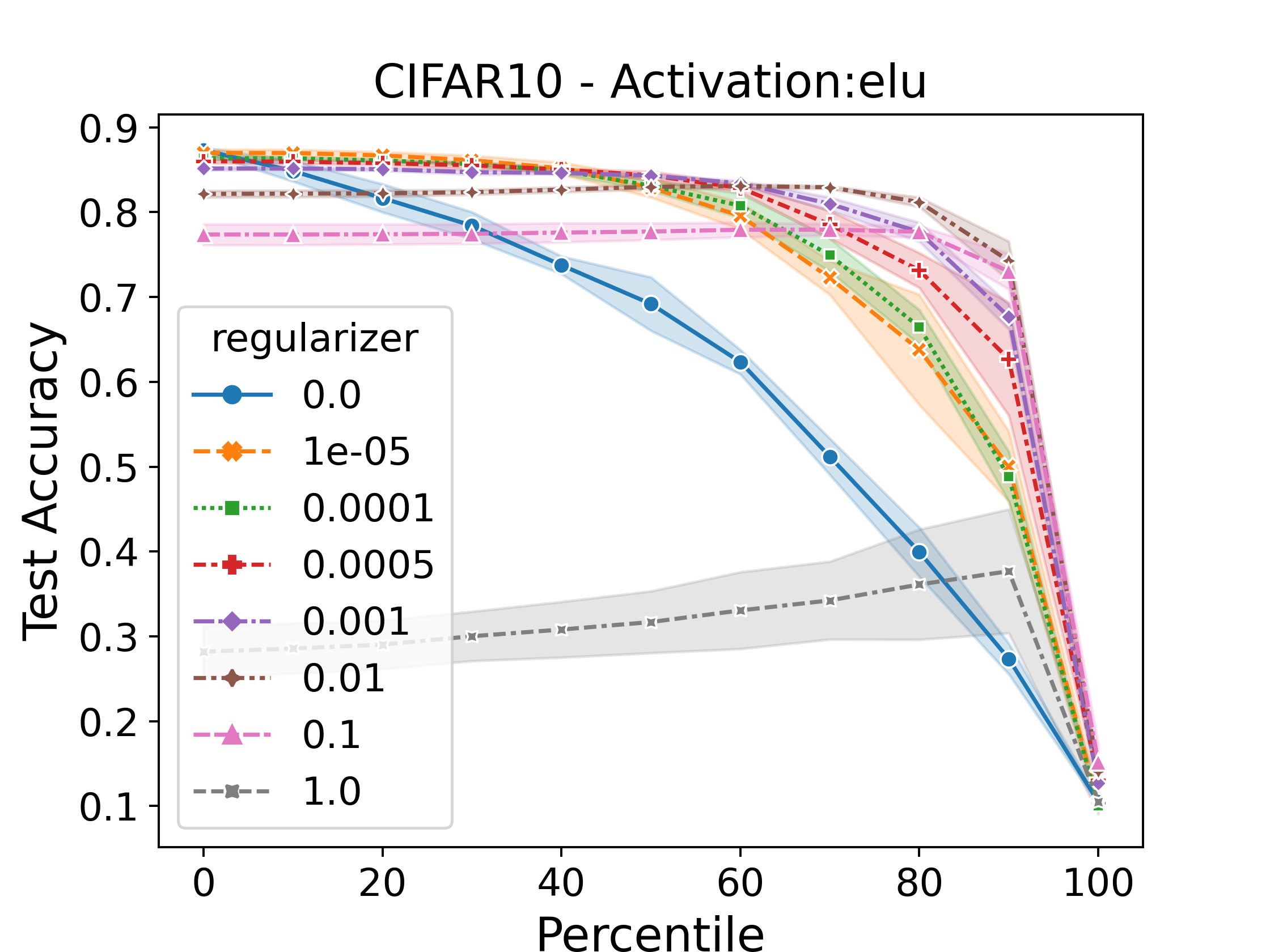}
		\caption{}
	\end{subfigure}
	\caption{Computed accuracy on the CIFAR10 dataset against the percentage of trimmed nodes (from the first of the two dense layers appended to the MobileNet-like architecture). The panels displays the performance of the network as according to each trimming procedure, and using weights ($W$) for the  $\ell_{1}$ regularizer. In panel (a) and (b) pre-training (based on two runs) and post-spectral filter, respectively; in panel (c) the reduction schem based on the absolute connectivity.}
	\label{fig:reg_plots}
\end{figure*}

To assess the flexibility of the schemes outlined  in Section III-B we here consider the CIFAR10 dataset and assume a modified MobileNetV2 \cite{Sandler_2018_CVPR} adding two dense layer at the end of the network. During training we freeze all the layers, except for the two appended dense layers. These latter are trained in the spectral domain. Working in this setting, the pruning is performed on the first dense layer by using strategies both (i) and (ii), as introduced in the main body of the paper. Here again the results are compared to those obtained when using the absolute value of the incoming connectivity as an alternative trimming criterion (see Figures \ref{fig:CIFARelu-multi}, \ref{fig:CIFARrelu-multi} and \ref{fig:CIFARtanh-multi}).  
As a further step in the analysis, we also introduce and test a $\ell_{1}$-norm regularization acting on the eigenvalues, so as to induce a sparse solution \cite{bach2012foundations}. All experiments are performed by using a MobileNetV2 based architecture. The first dense layer is made of 512 nodes with an ELU activation function (others activation functions yield analogous results). The following regularization loss functions are considered depending on whether the training takes place in the reciprocal (spectral layer) or direct space:
\begin{itemize}
	\item Spectral regularization $$L_r^{\text{spec}} = \gamma*\sum_{i=1}^{N_{\ell-1}} |\lambda^{(\ell-1)}_i|$$
	\item Connectivity regularization $$L_r^{\text{conn}} = \gamma*\sum_{i,j} |w^{(\ell-1)}_{ij}|$$
\end{itemize}
where $\gamma$ stands for a suitable regularizer weight.\\
Clearly $L_r^{\text{conn}}$ is equivalent to a regularization which acts on the incoming absolute connectivity. In fact,  $|\sum_i|x_i|| = \sum_i|x_i|$.\\ The $\ell_{1}$ regularization impacts significantly on the classification accuracy, as it can be clearly appreciated by direct inspection of Figure \ref{fig:reg_plots}.\\
Choosing the correct regularizer weight ($\gamma$), the performance of the network are stable across various range of pruning thresholds, even at the highest percentile.

\end{document}